\documentclass[preprint,12pt,number]{elsarticle}

\usepackage{amssymb}
\usepackage{amsmath}
\usepackage{lineno}
\journal{Nuclear Instruments and Methods A}

\begin{document}

\begin{frontmatter}



\title{Development of LGAD Sensor Technology on 12” Wafers for 3D Integration with advanced CMOS processes
} %% Article title

%% Author name
\author [Fermi]{Ron Lipton}
\author [SLAC]{Julie Segal}
\author [SLAC]{Angela Kok\fnref{fn3}}
\fntext[fn3]{Now at: Optronics Technology AS, Oslo,
Norway}
\author [Fermi]{Troy England\fnref{fn1}}
\fntext[fn1]{Now at: Emergence Quantum}
\author [Fermi]{Artur Apresyan}
\author [Fermi]{Davide Braga}
\author [Fermi]{Farah Fahim\fnref{fn2}}
\fntext[fn2]{Now at: Global Foundries}
\author [Fermi]{Shuoxing Wu}
\author [SLAC]{Christopher Kenney}
\author [SLAC]{Bojan Markovic}
\author [SLAC]{Ariel Gustavo Schwartzman}
\author [SLAC]{Lorenzo Rota}
%\author [SLAC]{Larry Ruckman}
\author [LLNL]{Arthur Carpenter}

\affiliation[Fermi]{organization={Fermilab},%Department and Organization
            addressline={P.O. Box 500}, 
            city={Batavia},
            postcode={60510}, 
            state={Illinois},
            country={USA}}
\affiliation[SLAC]{organization={SLAC National Accelerator Laboratory},
            addressline={2575 Sand Hill Road},
            city={Menlo Park},
            state={CA},
            postcode={94025},
            country={USA}}
\affiliation[LLNL]{organization={Lawrence Livermore National Laboratory},
            addressline={7000 East Avenue},
            city={Livermore},
            state={CA},
            postcode={94550},
            country={USA}}

%% Abstract
\begin{abstract}
We report on the design, simulation, and initial testing of Low Gain Avalanche Diode (LGAD) sensors fabricated in 65 nm CMOS technology on 12 inch wafers. These sensors were developed as part of a project to demonstrate full wafer to wafer hybrid bonding of sensors and 28 nm Readout Integrated Circuits (ROICs). The sensor submission and test included standard (reach-through) LGADs, AC coupled LGADs, and Deep Junction LGADs as well as various test structures. 
\end{abstract}


%%\item Research highlight 1
%%\item Research highlight 2

%% Keywords

\end{frontmatter}



\section{Introduction}
\label{intro}
%% Labels are used to cross-reference an item using \ref command.

Sensors and ROICs for particle detection and tracking have been utilizing integration and interconnection techniques established more than 55 years ago \cite{Miller1969}. Recent advances in interconnect technology such as hybrid wafer bonding and multi-tier stacking have the prospect to revolutionize capabilities of silicon-based detector systems. These 3D interconnect technologies offer low capacitance, heterogeneous integration, and very fine ($\geq   1 $ $\mu $m) interconnect pitch \cite{Lau2025} \cite{lipton2017}.

On the sensor side, emerging technologies such as LGADs, and new uses for older technologies such as SPADs and SIPMs provide  very fast time resolution in compact, low mass silicon-based sensors. A future intelligent tracking system may offer pixels that provide time, position, and angle information to a multi-tier readout system with stacked analog, digital and I/O layers to provide 5D tracking and on-sensor data filtering \cite{DSLGAD} \cite{7167712} \cite{turbiner20266dtrackingstudyusing} \cite{Yoo:2023lxy}.

\section{Project Goals}
The DOE-funded "3D Integrated Sensing Solutions" project is intended to demonstrate the incorporation of advanced technologies in particle and x-ray detectors by developing technology to achieve $10 \mu m$ spatial resolution, and $10 ps$ timing resolution in a 3D integrated package with low power consumption and high data throughput. The ultimate goal is to use 3D integration technology provided by Tower Semiconductor on 12" wafers to bond sensor wafers with 28 nm Readout Integrated Circuit (ROIC) wafers. We report on an initial phase of this project, the fabrication and test of fast sensors in the Tower 65 nm technology. The ROICs are described elsewhere \cite{Bragaiworid} \cite{MarkovicCPAD25}. 

\section{Sensors}
\begin{figure}[t]%% placement specifier
\centering%% For centre alignment of image.
\includegraphics[width=6.5cm,height=7.5cm]{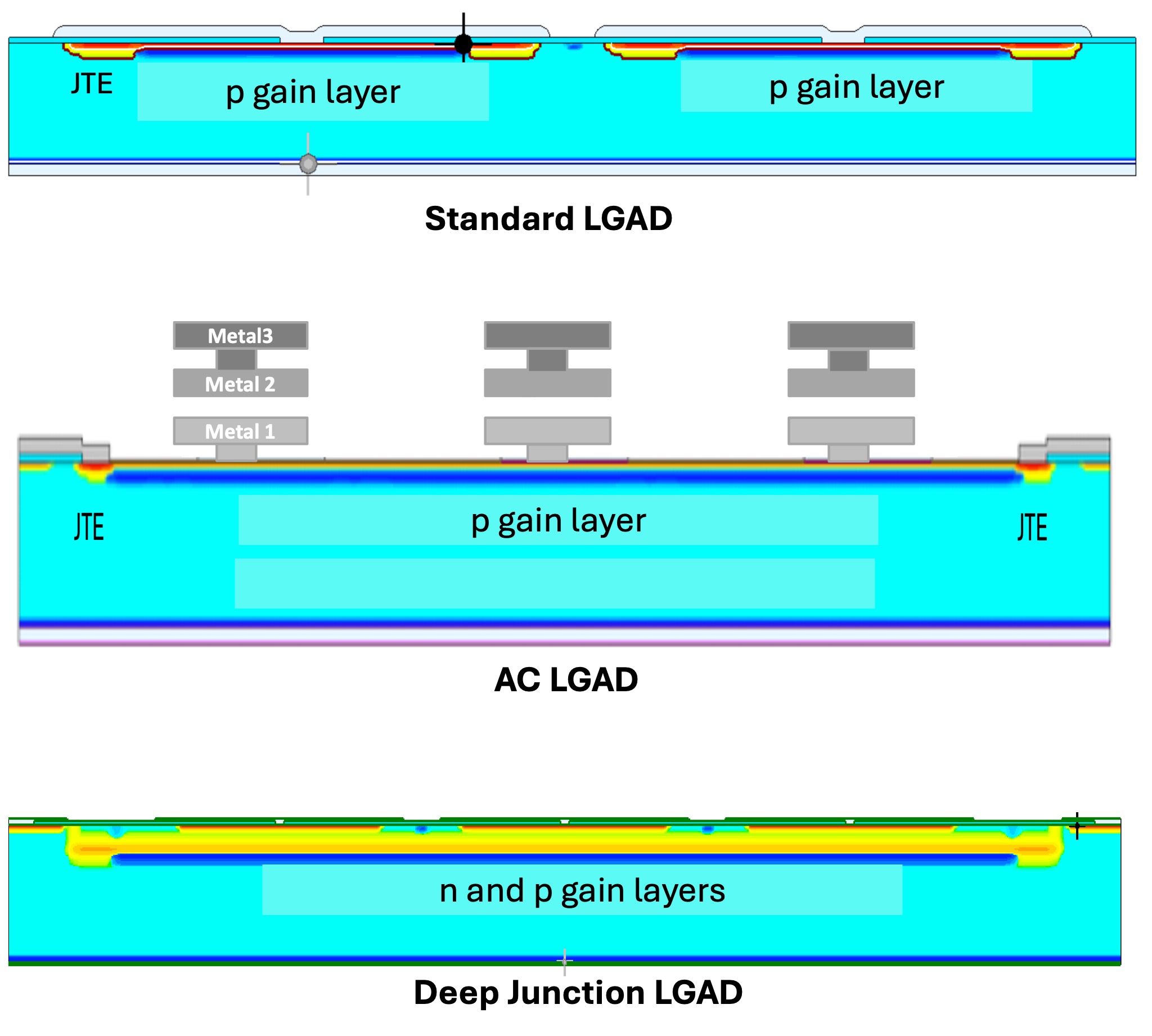}
\includegraphics[width=6.25cm]{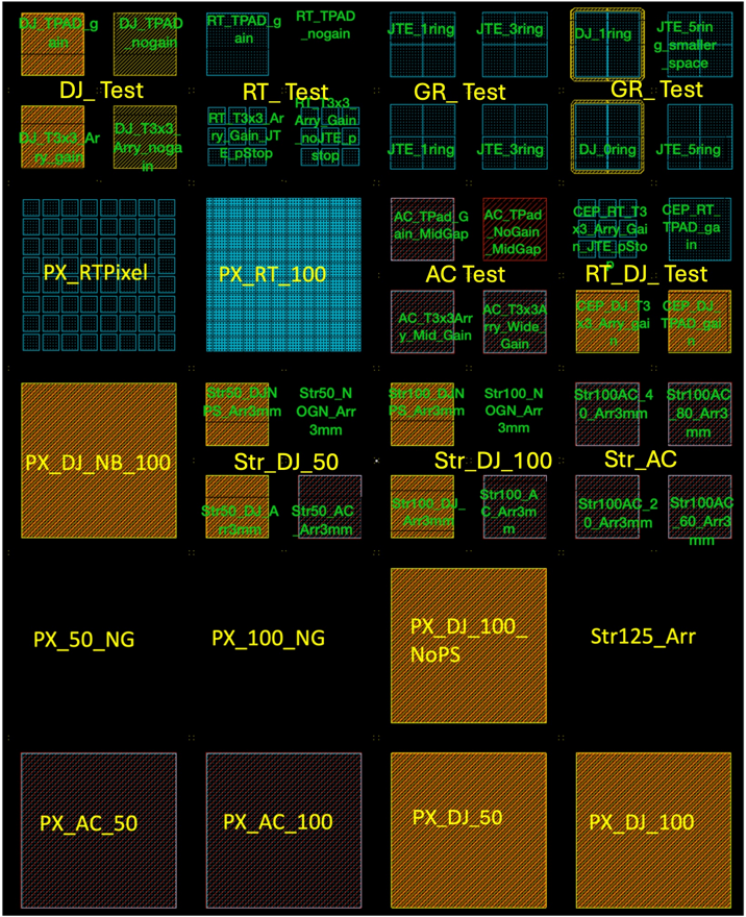}
%% Use \caption command for figure caption and label.
\caption{(left)Doping profiles for TCAD models of Standard(top) AC(Middle) and Deep Junction(bottom) LGADs. Red is n-type and blue is p-type. (right) Reticle layout for the Tower submission. The devices are contained in $6mm \times 6mm$ or $3mm \times  3mm$ cells.  Cells with orange backgrounds are DJ devices, blue background are standard (RT), Purple are AC and black have no gain layer.  PX are pixel cells, Str are strip test structures, and GR are guard ring test structures}
\label{fig1}
\end{figure}

The sensor wafer includes three types of Low Gain Avalanche Diodes (LGADs). These are (figure \ref{fig1}):
\begin{itemize}
    \item A "standard" LGAD \cite{Sadrozinski2018} \cite{Cartiglia2015} \cite{Sadrozinski2013}- a n+ electrode with a "reach through" high energy boron implant that defines a high field gain region between the shallow n+ and deep p implants.  The resulting high electric field is terminated by a Junction Termination Extension (JTE) deep n implant to avoid edge breakdown. This design has the disadvantage that the area taken up by the JTE reduces the effective fill factor and drives the pixel size.
    \item AC Coupled LGAD - In this variant the basic geometry remains the same, but the output signal is AC coupled through a surface dielectric to the readout pixel electrode. A resistive cathode implant is used to develop the signal voltage and limit charge spread \cite{Giacomini2019} \cite{MANDURRINO2020163479}.  This design requires only a single JTE structure that surrounds the entire pixel array rather than one per pixel.  The charge sharing provided by the resistive cathode can provide excellent position resolution \cite{Heller2022}. The need for AC signal processing and a coupling capacitor can complicate the system design.
    \item Deep Junction LGAD - This is a newer variant which utilizes buried implants to define the gain layer \cite{ayyoub2021} \cite{Apresyan:2021tkb}. The gain field is confined to the bulk of the semiconductor, eliminating breakdown near the pixel edges so the pixel pitch is not constrained by fill factor. Previous studies have used wafer bonding or graded epitaxial layers \cite{ayyoub2021}. In our version we use high energy (MeV) implants to form the buried gain layers. The Tower process provides sufficient energy for the deep high energy buried implants.

\end{itemize}

The reticle included $6mm \times 6mm$ 50 and 100 micron pitch AC, Deep Junction and "no gain" pixel arrays designed to bond to the 28 nm ROIC. Test structures were placed in $3mm \times 3mm$ sub cells. These included pixel arrays of each device type suitable for bench testing as well as guard ring and strip structures. Each design was surrounded by a multi guard ring structure.

%% Use \subsection commands to start a subsection.

\section{Process Flow and Constraints}
The process flow is based on a modified Tower $65 nm$ process \cite{10.1117/12.3013899} with a $10 \mu m$ thick epitaxial layer and three metal routing layers plus top metal. We repurposed existing process layers to conform with the Tower process flow. We were allowed to modify the implant doses and energies to optimize the LGAD designs however the basic process flow and annealing schedule were not modified. 

The first layer oxide thickness usually used for the AC coupling dielectric is defined by the standard Tower process. This oxide is thicker than the metal one-metal two oxide. Therefore we opted to utilize the capacitance between metal one and metal two for the AC LGAD coupling capacitance rather than the usual metal one to substrate capacitance. The capacitor area is limited by the metal density design rules. Alternative techniques, such as Metal-Insulator-Metal (MIM) capacitors or finger capacitors may be used in future runs.

The 65 nm CMOS process has much more complex design rules than a typical sensor process. Our existing sensor designs had to be adapted to comply with metal width, spacing and density design rules. There were a number of irrelevant errors due to the unusual sensor-only design during the design rule checks, requiring an extensive period of review and modification. Our development run included 12 wafers with varying doping and energy splits. An additional run is underway with additional variants based on initial testing of the current batch.

\section{TCAD Simulation}
\begin{figure}[t]%% placement specifier
\centering%% For centre alignment of image.
\includegraphics[width=7.5cm ]{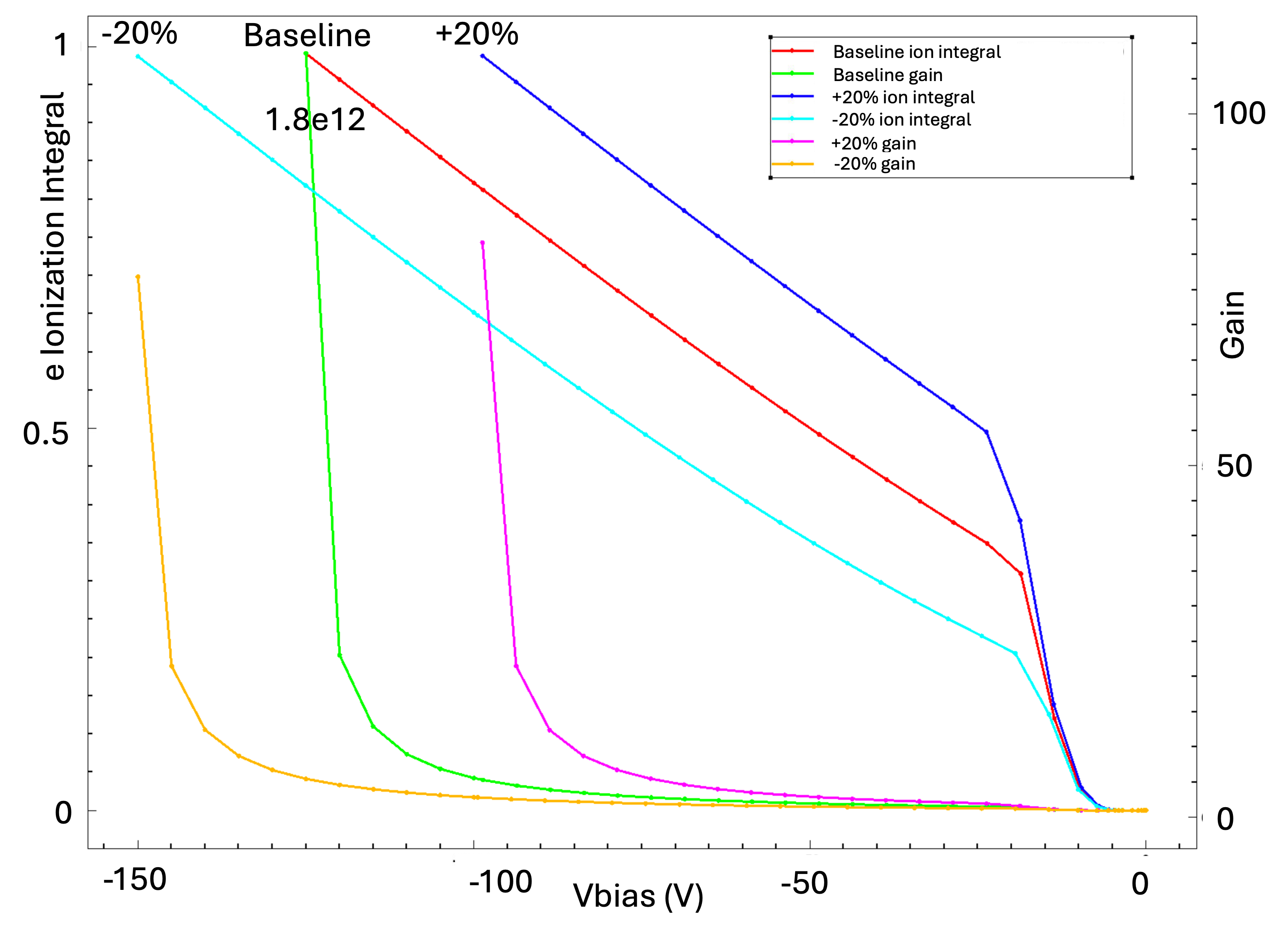}
\includegraphics[width=6cm]{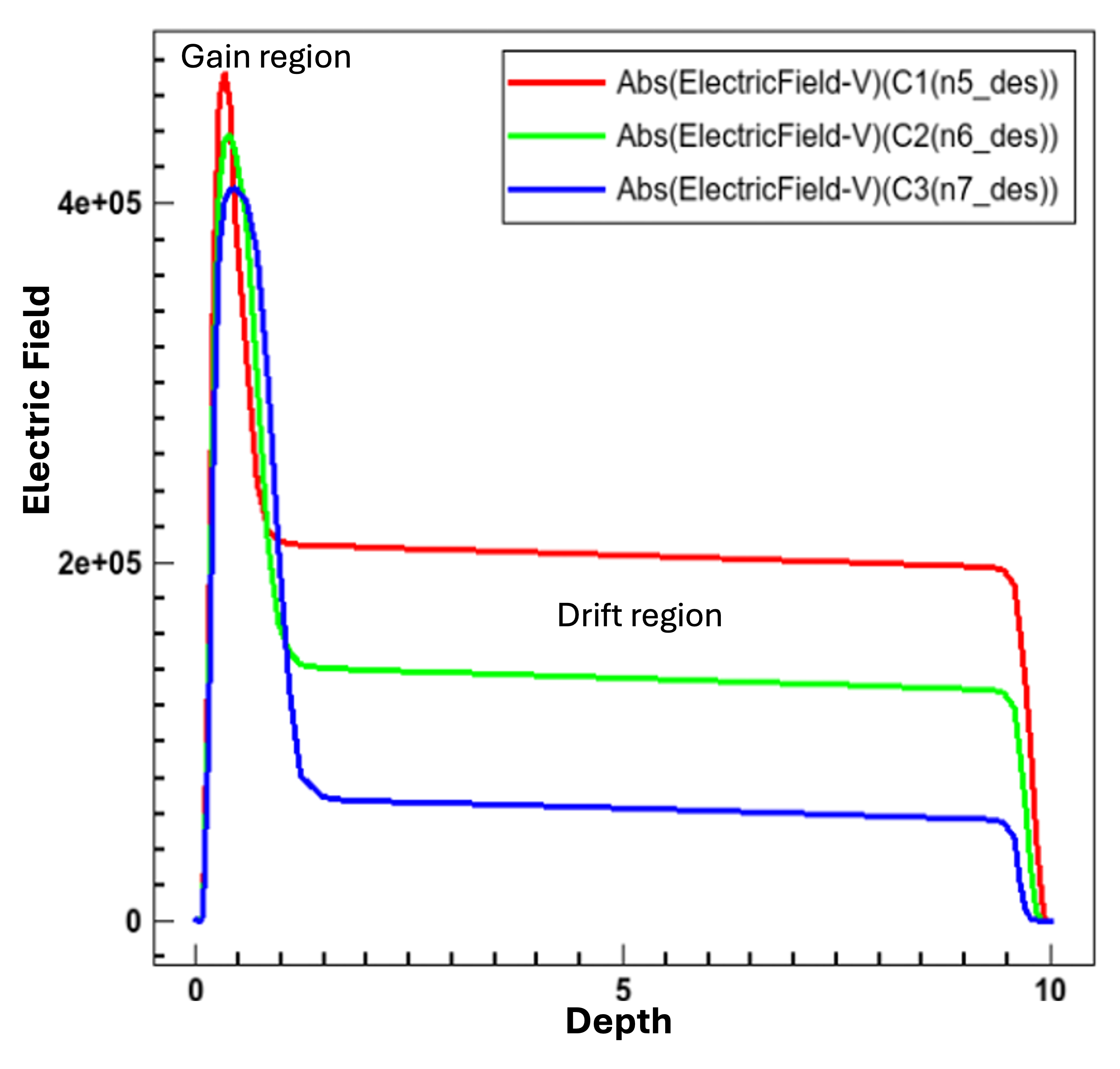}
%% Use \caption command for figure caption and label.
\caption{(left) Plot of ionization integral and corresponding calculated gain vs bias voltage for standard LGADs with boron implant doses varied by 20\%. (right) Plot showing the associated gain region and drift fields. }
\label{TCAD}
\end{figure}

TCAD simulations were used to develop the baseline and process split variations.  Our devices were simulated using a full process flow Synopsys Sentaurus simulation \cite{synopsys}.  Ionization integrals and leakage currents were used to establish  projected operating voltages. We also studied the drift and gain fields in various regions, overall gain, and pulse shapes. Figure \ref{TCAD} shows some representative results.

The Ionization Integral (II) is particularly useful as it is related to the device gain ($\sim \frac{1}{1-II}$) and is strongly dependent on the electric field. As the device is biased, the electric field is initially confined to the region between the deep implant(s) and the cathode, corresponding to a 1-3 micron depth. The initial large ionization integral slopes in figure \ref{TCAD} correspond to the rapid variation in the electric field in this small depleted depth before depletion of the gain layer(s) ($\delta E \approx \delta V_{bias} / Depletion\  depth$).
After full depletion the lower slope is characteristic of the full thickness of the device.

The thin Tower epitaxial layer implies that the gain of these devices are more sensitive to bias voltage than thicker devices.

%Fig \ref{TCAD}
Simulations were performed to breakdown and the 0.95 and 0.9 ionization integral points were used as gain reference points. For each LGAD variant we defined a "baseline" set of bias voltage values as well as higher and lower operating point variants.  
For standard devices of a given gain layer doping the operating voltage is reduced with increasing implant energy or higher gain layer doping. The gain layer doping dependence is due to the increased value of the ionization integral at the inflection point of the  II vs $V_{bias}$ curve (figure \ref{TCAD}) where the gain layer is depleted. 

The AC variants used the same dose and energy parameters as the standard devices with lower cathode implant doses to allow for charge sharing and coupling to the AC capacitance.  A simple SPICE 10x10 pixel model was used with the simulated sheet resistances in collaboration with the ASIC designers. A more complex Cadence Virtuoso model was used to estimate charge sharing and pulse shapes in a 100x100 pixel array. A 3D TCAD simulation was constructed to further understand the pulse shapes and confirm the SPICE and Cadence modeling. 

Deep Junction (DJ) devices are more complex since the gain field occupies the region between high energy implants, which may also have significant overlap.  We observe a 5\% variation in the gain field in the inter-pixel region, which affects the gain.  In this case both the deep boron and shallower phosphorus implant energies were varied but the doses were kept constant.

A number of simulations were performed to optimize the multiple guard ring structure. The baseline structure has 9 n-rings with JTE layers and 2 $\mu m$ spacing.  There are also floating p -rings in case of possible surface inversion.  

%\begin{figure}[t]%% placement specifier
%\centering%% For centre alignment of image.
%\includegraphics[width=10cm]{reticle.png}
%% Use \caption command for figure caption and label.
%\caption{Reticle layout for the Tower submission. The devices are contained in $6mm \times x 6mm$ or $3mm \times  3mm$ cells.  %Cells with orange backgrounds are DJ devices, blue background are standard, Purple are AC and black have no gain layer.}
%\label{reticle}
%\end{figure}

Test structures with smaller LGADs and diodes as well as several guard ring varieties were included for initial testing and evaluation of layout variants.  

\section{Testing}
\begin{figure}[t]
\centering
\includegraphics[width=12cm]{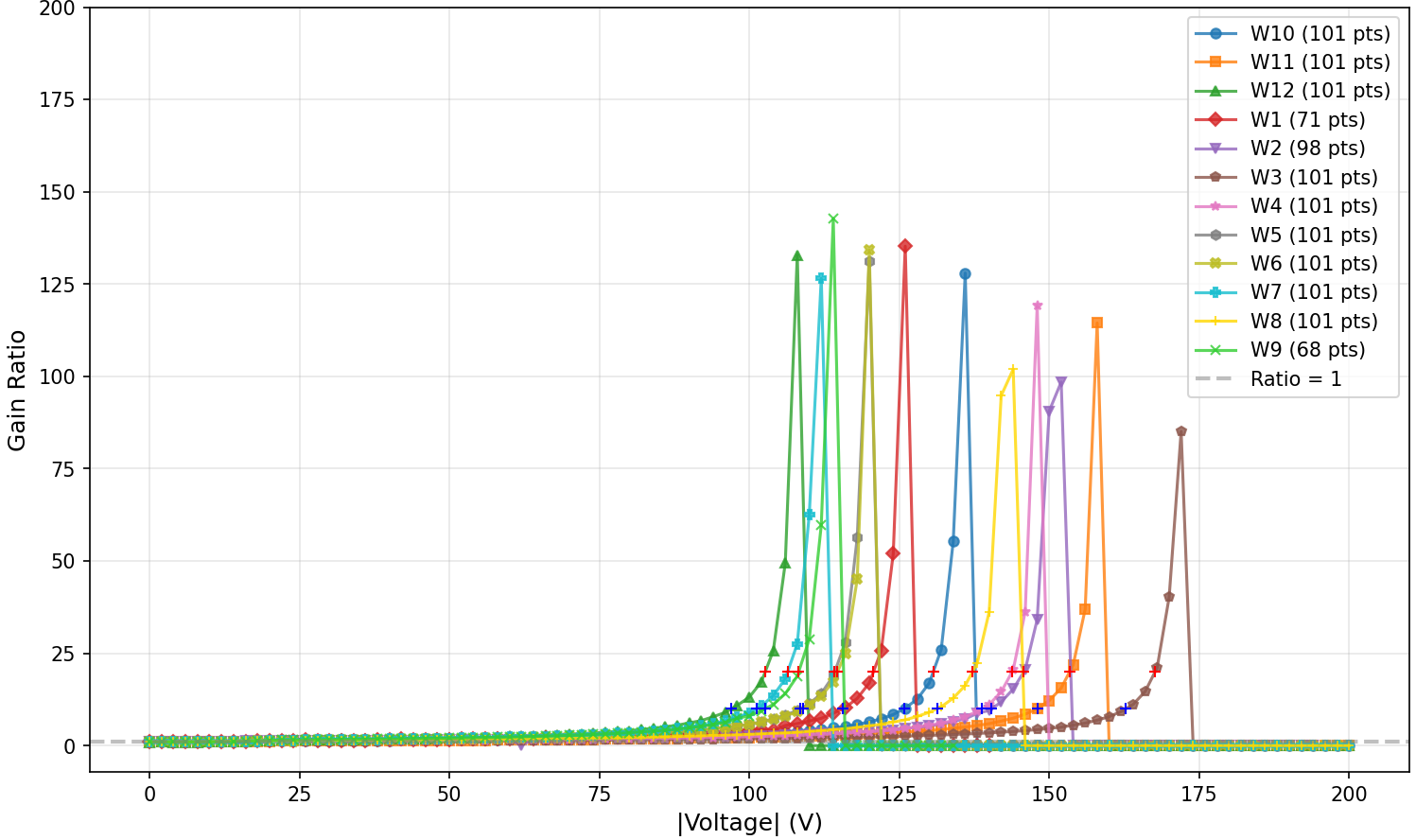}
\caption{660nm LED measurements of AC LGADs in wafers 1-12. Gain is defined by comparing test structures with and without gain layers as $(Current(LED_{on})-Current(dark))_{gain}/(Current(LED_{on})-Current(dark))_{nogain}$ Small crosses show the gain ratio of 10 (blue crosses) and  20 (red crosses) operating points.}
\label{ACgain}
\end{figure}

Wafers were received in February 2026. We report here on the first phase of tests. Our initial tests compared the voltage-current characteristics observed  in the no gain reference diode to AC, standard, and deep junction test structures under 660 nm LED illumination and dark conditions (figure \ref{ACgain}). These measurements established values for gain vs bias as well as breakdown voltage with and without illumination.  A 1064 nm laser as well as a Ruthenium beta test stand were used to examine pulse response and timing. Beam tests were also recently performed at the H6 120 GeV test beam at CERN. 

Two wafers were sent for Secondary Ion Mass Spectroscopy (SIMS) profiles of the gain and cathode implants. The measured profiles are consistent with those expected from TCAD. Radiation tests are planned for the National Ignition Facility in at Lawrence Livermore National Laboratory and at the University of California, Davis. 

\subsection{AC and standard LGADs}
The AC and standard LGADs share the same gain layer implants. Figure \ref{ACgain} shows the LED results from the AC LGADs showing a range of gain of 20 operating voltage between 116 and 160 volts. In general, the TCAD predictions for the x10 and x20 gain values for the AC LGADs are within 10 Volts of the LED measurements. The lower dose cathode implant dose in the AC LGAD does affect the  operating point, resulting in a  standard LGAD operating voltages between 20 and 30 volts higher than the value for the AC LGADs with the same gain layer dose and energy.

\begin{figure}[t]
\centering
\includegraphics[width=12.5cm]{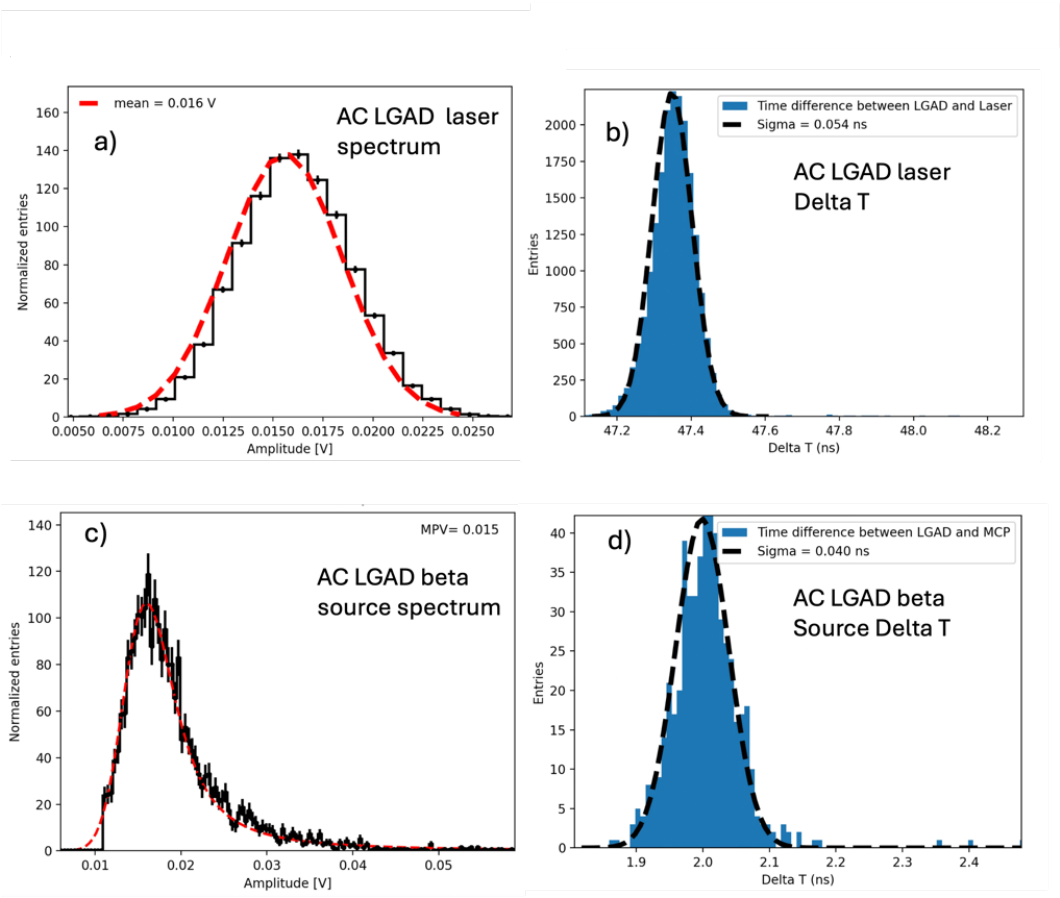}
%% Use \caption command for figure caption and label.
\caption{Laser (a,b) and beta source (c, d) measurements of AC LGADs. The beta source shows a typical Landau distribution, the laser pulse has to penetrate several dense metal layers.  The arrival time resolutions are similar, with the laser resolution at $54 ps$ and the beta source at $40ps$.}
\label{AC_laser}
\end{figure}

\begin{figure}[t]
\centering
\includegraphics[width=8cm]{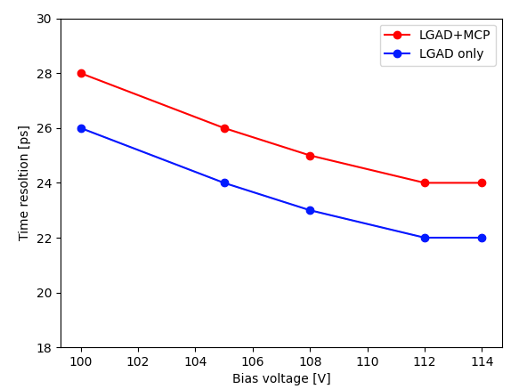}
%% Use \caption command for figure caption and label.
\caption{Time resolution vs bias voltage for a $0.66 \times 0.66$ mm AC LGAD tested using a beta source with a microchannel plate (MCP) timing reference}
\label{DT}
\end{figure}

Figure \ref{AC_laser} shows results for laser and beta tests for a 2x2 mm AC LGAD. The time resolutions in this test are $54 ps$ and $40 ps$ respectively.  Bench tests of a 0.66 x 0.6 mm LGAD are shown in figure \ref{DT}. In this case the time resolution near breakdown improves to $23 ps$. A similar value for time resolution was observed in a recent beam test.

\subsection{Deep Junction}
The Deep Junction devices are more unstable than the AC or standard LGADs, with a slow drift of the breakdown voltage with time and sharp breakdown characteristics. This makes detailed measurements difficult. We do see gain and good time resolution in some of the test structures in laser, beta and beam tests.  Deep junction implants in the 12 wafers currently in process were modified from the current set of 12 wafers to increase the spacing between boron and phosphorus layers. 

\subsection{Continuing Work}
Strip structures have not yet been tested, these should yield charge sharing information in the AC LGADs as well as measurements of the coupling capacitance values. Pixel arrays are being bonded, which should provide information on the performance of the 28 nm ROICs with AC and DJ LGADs.  Irradiation and beam tests are also ongoing.  Finally, a set of 12 wafers should be delivered soon with additional doping variants.

\section{Conclusions}
We have designed, produced, and begun the testing of AC, reach-through and deep junction LGADs fabricated in a 65 nm 12” wafer CMOS process. The process utilized a standard CMOS process with  modified implantation energies and doses. This is the first fabrication of a deep junction LGAD  utilizing high energy implants to form the gain layers.  Our initial tests indicate that all LGAD types give adequate gain and good time resolution.  Pixel integration with the 28nm readout chips and more detailed characterization of the sensor properties is underway.

These sensors were designed with the intention of eventual hybrid (3D) bonding of full wafers with 28nm readout wafers in a second phase of the project.  Availability of these technologies can provide very fine pitch bonding ($1-3 \mu m$), low capacitance, separate optimization of sensor and ROIC technologies,  radiation hardness, and multi-tier assembly of ROICs to include complex signal processing for future generations of experiments. 

\section*{Acknowledgments}
The 3D Integrated Sensing Solutions Collaboration was supported by DOE Office of High Energy Physics FWP 101062. Fermi National Accelerator Laboratory (Fermilab), is a U.S. 
Department of Energy, Office of Science, Office of High Energy Physics HEP User Facility. Fermilab is managed 
by Fermi Forward Discovery Group, LLC, acting under Contract No. 89243024CSC000002. 
Work at SLAC National 
Accelerator Laboratory is supported by the U.S. Department 
of Energy, Office of Science under contract number DE-AC02-76SF00515. 
This work was performed under the auspices of the U.S. Department of Energy by Lawrence Livermore National Laboratory under Contract DE-AC52-07NA27344.

%%       \citet{<label>} ==> Lamport (1994)
%%       \citep{<label>} ==> (Lamport, 1994)
%%

%\bibliographystyle{plain}
\bibliographystyle{elsarticle-num}
\bibliography{Ultima_LGAD}
\end{document}